# THE EFFECT OF GRAVITATIONAL SCATTERING ON THE ANISOTROPY OF THE COSMIC BACKGROUND RADIATION

Toshiyuki Fukushige, Junichiro Makino[†] and Toshikazu Ebisuzaki

Department of Earth Science and Astronomy,
[†]Department of Information Science and Graphics,
College of Arts and Sciences, University of Tokyo,
3-8-1 Komaba, Meguro-ku, Tokyo 153, Japan
E-mail:fukushig@chianti.c.u-tokyo.ac.jp

## Abstract

The homogeneity of the cosmic microwave background radiation (CBR) is one of the most severe constraint for theories of the structure formation in the universe. We investigated the effect of the gravitational scattering (lensing) of galaxies, clusters of galaxies, and superclusters on the anisotropy of the CBR by numerical simulations. Although this effect was thought to be unimportant, we found that the gravitational scatterings by superclusters can significantly reduce the anisotropy of the CBR. We took into account the exponential growth of the distance between two rays due to multiple scatterings. The bending angle of each ray grows through the random walk process. On the other hand, difference between two rays grows exponentially while it is small. This exponential growth is caused by coherent scatterings that two rays suffer, and was neglected in the previous studies. The gravitational scattering by superclusters reduces the observed temperature anisotropy of the CBR at present time approximately by 40–60 % from that at the recombination time for angular scale up to a few degrees if the supercluster were formed at $z = 2$–$4$.

Subject headings: cosmic microwave background, gravitational lensing, large-scale structure of Universe

## 1. Introduction

Observations have shown that the cosmic microwave background radiation (CBR) is extremely isotropic. Observed upper limits of the temperature fluctuation, $\Delta T/T$, of the CBR are $\sim 4.5 \times 10^{-5}$ at $4'.5$ (Uson and Wilkinson 1984), and $\sim 2.1 \times 10^{-5}$ at $7'.15$ (Readhead et al. 1989). Recent observation with COBE showed that the temperature fluctuation is $(1.1 \pm 0.18) \times 10^{-5}$ at 10 degrees (Smoot et al. 1992), and the observation at South Pole showed that the upper bound is $1.4 \times 10^{-5}$ at a degree scale (Gaier et al. 1992). If the density fluctuation is actually small as suggested by these observations, the structure formation models which are consistent with the "observed" temperature anisotropy are rather few. For example, the baryon dominant model is ruled out (Peebles and Silk 1990;

Gouda, Sasaki and Suto 1989). In addition, the standard CDM model is only marginally acceptable (Gouda and Sugiyama 1992).

During its travel from the last scattering surface, the CBR is gravitationally scattered by astronomical objects, such as galaxies, clusters of galaxies, and superclusters. We investigated the effect of the gravitational scattering on the anisotropy of the CBR, taking into account the exponential growth of the distance between nearby rays through multiple scatterings. We found that the temperature anisotropy of the CBR can be reduced by 40–60% through scatterings by superclusters.

The smoothing of the CBR by the gravitational lensing has been investigated by many researchers (Blanchard and Schneider 1987; Kashlinsky 1988; Tomita 1988; Cole and Efstathiou 1989; Sasaki 1989; Linder 1990; Tomita and Watanabe 1989; Watanabe and Tomita 1991). Sasaki (1989) and Linder (1990) gave the mathematical formula of the gravitational lensing on the angular correlation function of the temperature anisotropy.

Kashlinsky (1988) derived equations that describe the multiple gravitational lensing of the CBR, and concluded that the original fluctuation was smoothed out on scales up to several arcminutes. Cole and Efstathiou (1989) pointed out that Kashlinsky overestimated this effect because he modeled galaxy clusters as point masses. They argued that the effect is negligible using the formula of increase of the beam width derived by Gunn (1967). Blanchard and Schneider (1987) and Watanabe and Tomita (1991) also obtained similar results. Tomita and Watanabe (1989) performed numerical simulations of propagation of light, and concluded that the effect of the gravitational lensing by clusters of galaxies was small.

Gunn (1967) derived the formula for change of the beam width under an assumption that total effect of increase of the beam width is expressed by a superposition of that of small scatterings. However, this assumption is not appropriate. If the distance between two rays is small, they are scattered coherently. The average increase of the distance between rays by one scattering is proportional to its width, since the difference in the deflection angle is caused by the tidal force. As a result, the distance increases exponentially by multiple scatterings. This exponential growth continues until the scattering becomes incoherent. Therefore, evidently, the increase of the distance between rays cannot be expressed by the simple superposition.

In the field of stellar dynamics, the exponential instability of a small difference of initial condition has been well known (Miller 1964; Lecar 1968; Sakagami and Gouda 1990; Suto 1991; Kandrup and Smith 1991; Kandrup, Smith and Willmes 1992; Quinlan and Tremaine 1992; Huang, Dubinski and Carlberg 1993). However, a clear theoretical understanding of this exponential growth was given only recently by Goodman, Heggie and Hut (1993).

Therefore, it is not surprising that this effect has been neglected in the study of the effect of the gravitational scattering on the CBR.

In this letter, we investigate the exponential growth of the distance between nearby rays by numerical calculations, and evaluate its effect on the anisotropy of the CBR. We calculated sets of path of two photons in a uniform distribution of the scattering objects. Our calculation showed that the angle between two rays increases exponentially up to an angle $\theta_{\mathrm{pr}}$ given by $\theta_{\mathrm{pr}} = \sqrt{4\pi} N^{-\frac{1}{2}}$, where $N$ is the number of the scattering objects within the horizon of the present universe in the case of a flat universe. When the angle becomes larger than $\theta_{\mathrm{pr}}$, it increases in proportion to $t^{1/2}$. This is because the scattering becomes incoherent between rays.

The angle grows exponentially only when the half-mass radius of the scattering objects, $r_{\mathrm{h}}$, is smaller than the projected mean particle distance: $d_{\mathrm{pr}} = \sqrt{4\pi} N^{-\frac{1}{2}} R_{\mathrm{H}}$, where $R_{\mathrm{H}}$ is the distance to the horizon. We found that superclusters satisfies this condition of the exponential growth, $r_{\mathrm{h}} < d_{\mathrm{pr}}$, and that $\theta_{\mathrm{pr}}$ for supercluster is a few degrees.

We conclude that the anisotropy of the CBR is smoothed up to scale of a few degrees. We evaluated the angular correlation function using the formula derived by Wilson and Silk (1981), and Sasaki (1989). We found that the gravitational scattering due to superclusters can decrease the anisotropy of the CBR approximately by 40–60% if the beam width of antenna is not much smaller than the intrinsic angular scale of the fluctuation.

## 2. Numerical Calculations

We calculated sets of paths of two photons in a uniform and random distribution of the scattering objects. We used the post-Newtonian equations of motion for photon (Will 1981):

$$\frac{d^2\mathbf{x}}{dt^2} = 2\nabla\phi - 4\mathbf{n}(\mathbf{n}\cdot\nabla\phi), \tag{1}$$

where $\mathbf{x}$ is the position of photon, $\phi$ is the gravitational potential, and $\mathbf{n}$ is the initial direction vector which satisfies $|\mathbf{n}| = 1$. This post-Newtonian approximation is always valid in a real universe, unless we consider a universe dominated by supermassive black holes. The maximum deflection angle of a scattering by an object is determined by the depth of its potential well. The depth of the potential well of galaxies, clusters of galaxies or superclusters is by far smaller than $c^2$, where $c$ is the light velocity. In the numerical calculations, we ignored the second term of the right-hand side of equation (1), since our purpose is to obtain the trajectory of a photon. The second term changes the deflection angle through the change of the light velocity. Since the changes of the velocity and the deflection are both $O(\phi/c^2)$, the contribution of the second term to the deflection is $O[(\phi/c^2)^2]$, which is negligible.

We calculated the trajectory of a photon using the gravitational potential expressed as

$$\nabla \phi = \sum_j^N \frac{GM_{r,j}(r_{s,j})}{r_{s,j}^3}\mathbf{r}_{s,j} + \frac{GM}{R^3}\mathbf{x}, \qquad (2)$$

where $\mathbf{r}_{s,j} = \mathbf{x}_{s,j} - \mathbf{x}$, $G$ is the gravitational constant, $N$ and $M$ are the number and the total mass of the scattering objects, $\mathbf{x}_{s,j}$ and $m_{s,j}$ are the position and mass of the $j$-th scattering object, $R$ is the radius of the sphere in which the scattering objects are uniformly distributed, and $M_{r,j}(r)$ is the mass contained within a sphere of radius $r$ and $M_{r,j}(r=\infty) = m_{s,j}$. We used the system of units in which $G = R = c = 1$ and $M = 1/2$. If we specify this sphere as a flat universe, the unit of time is $\sqrt{2}/H_0$ and the radius $R$ is $\sqrt{2}c/H_0$, where $H_0$ is the Hubble constant. The masses of the scattering objects were set to be equal to each other, i.e., $m_{s,j} = M/N$. The second term of the right-hand side of equation (2) was introduced to cancel the global harmonic potential due to the scattering objects. We started the numerical integration of the trajectory of photons at the surface of the sphere. The initial velocity vector, $\mathbf{v}_0$, of a photon points to the center of the sphere and $|\mathbf{v}_0| = 1$. To calculate the growth of the distance between initially nearby rays, we calculated the trajectory of two photons from slightly different initial conditions. We performed calculations for three cases: $N = 128, 1024, 8192$. We used GRAPE-2A (Ito et al. 1993), a special-purpose computer for $N$-body problem, for the calculation of the gravitational force. We integrated the orbits with the 4-th order Runge-Kutta scheme with an automatic timestep adjustment (Press et al. 1986) and the Hermite scheme (Makino and Aarseth 1992). The details of our calculation are presented elsewhere (Fukushige et al. 1994).

In figure 1, growth factor, $\alpha_{1.5} \equiv \theta_{1.5}/\theta_0$, is plotted against $\theta_0/\theta_{pr}$ where angles $\theta_0$ and $\theta_{1.5}$ are the median values of 200 orbit pairs of $\theta$ at $t = 0$ and $1.5$, respectively. In these calculations, the scattering objects were set to be point mass (i.e. $M_{r,j}(r > 0) = m_{s,j}$). Figure 1 indicates that the behavior of $\alpha_{1.5}$ is divided into three cases, depending on the value of $\theta_0/\theta_{pr}$, i.e., exponential, transitional, and diffusive region. If $\theta_0/\theta_{pr} < 0.1$, the growth factor, $\alpha$, is about 15–20, independent of the initial condition. In this region, the angle grows exponentially. If $0.1 < \theta_0/\theta_{pr} < 1$, the growth factor $\alpha_{1.5}$ is roughly represented by $\theta_{pr}/\theta_0$. In this region, $\theta$ increases up to $\theta_{pr}$. If $\theta_0/\theta_{pr} > 1$, the growth factor, $\alpha_{1.5}$, is of the order of unity. These behaviors are independent of the number of the scattering objects, $N$.

The angle grows exponentially only when the half-mass radius of the scattering objects, $r_h$, is smaller than the projected mean particle distance, $d_{pr}$. We calculated the orbits with different mass distributions for scattering objects in order to investigate the effect of the size. In figure 2 the growth factor is plotted against $r_h/d_{pr}$ for the King model (King 1966) with dimensionless central potential $W_0 = 5, 8, 12$ (where the concentration parameter are $\log[r_t/r_c] = 1.0, 1.8, 2.7$ for $W_0 = 5, 8, 12$, respectively, where $r_t$ is tidal radius and $r_c$ is core radius) and the Plummer model (Plummer 1911). The number of particle, $N$, of scattering objects was 1024 and the initial angle was chosen so that $\theta_0/\theta_{pr} \ll 1$. The behavior is almost independent of the softening law. The slope of decrease of the growth factor for larger softening depends on the mass distribution, and is a little smaller for more concentrated distribution.

## 3. Smoothing of Temperature Anisotropy of the CBR

In this section, firstly, we derive a formula for the growth factor, $\alpha$, in the expanding universe. Next, we estimate the anisotropy of the CBR in term of the angular correlation function. Here, we assume that the universe is flat ($\Omega$=1) for simplicity.

The growth factor, $\alpha_{LS}$, of the angle between two photons that come from the last scattering surface is calculated as $\alpha_{LS} = [w(z_F) + \theta(z_F)R_{FL}]/(\theta_0 R_{0L})$, where $w$ and $\theta$ are the distance and angle between photons at given redshift $z$, respectively, $\theta_0 = \theta(z=0)$, $z_F$ is the redshift when the scattering object were formed, and $R_{0L}$ and $R_{FL}$ are the distances to the last scattering surface from $z = 0$ and $z = z_F$, respectively. The relation among the above parameters are illustrated in figure 3.

The distance $w$ and angle $\theta$ are approximately given by the equations:

$$\frac{dw}{dt} = -c\theta_0 - \frac{w}{t_e}, \qquad w|_{z=0} = 0, \quad \text{and} \quad c\theta = -\frac{dw}{dt}. \tag{3}$$

The e-folding time, $t_e$, is given by $t_e = \tau/\sqrt{G\rho}$ (Goodman, Heggie and Hut 1993), where $\rho$ is the mass density of the scattering object. Using the result of our calculations, $\tau$ is given by $\tau = 3\sqrt{3}(4\sqrt{2\pi}\ln\alpha_{1.5})^{-1}$. For the case that the scattering object is point mass, $\tau \sim 0.18$. Solving the equations (3), we obtain the growth factor $\alpha_{LS}$:

$$\alpha_{LS} = (1 + z_F)^{\frac{3}{2}\eta}, \tag{4}$$

where $\eta = \Omega_s^{\frac{1}{2}}(6\pi)^{-\frac{1}{2}}\tau^{-1}$, and $\Omega_s$ is the density parameter of the scattering object.

In Table 1, we summarize estimates of $r_h/d_{pr}$ and $\theta_{pr}$ for galaxies, clusters of galaxies, and superclusters. The projected mean particle distance, $d_{pr}$, is calculated as $d_{pr} = (3d_{sep})^{\frac{1}{2}}R_H^{-\frac{1}{2}}\Omega_s^{-\frac{1}{4}}d_{sep}$. The time scale of the exponential growth, $t_e$, is scaled by crossing time, $t_{cr}$. Since the radius of the sphere to which the scattering objects are projected is $\sim ct_{cr}$, it becomes a function of the density parameter, $\Omega_s$. Therefore, the distance $d_{pr}$ is also a function of $\Omega_s$. The details are discussed in Goodman, Heggie, and Hut (1993). We adopt $R_{\rm H}(=2c/3H_0)$ of $2h^{-1}$Gpc, where $h$ is the Hubble constant in unit of 100kms$^{-1}$/Mpc.

Using figure 3, we estimate the growth factor, $\alpha_{\rm LS}$, and show then in Table 1. The gravitational scatterings by superclusters increase the angle by a large factor. Here, we assume that the superclusters have a large fraction of the mass in the universe. This can be justified from the observation showing that the universe consists of the void with very low density and the structures of the scale of supercluster. We adopted the size of supercluster of $10h^{-1}$Mpc. The superclusters typically have elongated shapes. The effect of the gravitational scattering is determined by the length of the shortest axis. It is unclear whether galaxies ($\Omega_{\rm s} \sim 0.2$) and clusters of galaxies ($\Omega_{\rm s} \sim 0.1$) have significant effect or not.

In the following, we give a quantitative estimate of the anisotropy of the CBR in terms of the angular correlation function using the formula derived by Wilson and Silk (1981) and Sasaki (1989). Temperature fluctuation of the CBR averaged over the beam pattern of an antenna is given by

$$\left\langle \left(\frac{\delta T}{T}(\theta;\sigma)\right)^2 \right\rangle \equiv 2[C(0;\sigma) - C(\theta;\sigma)], \tag{5}$$

where

$$C(\theta;\sigma) \simeq \frac{1}{2\sigma^2}\int_0^\infty \varphi C(\varphi) \exp\left[-\frac{1}{4\sigma^2}(\theta^2+\varphi^2)\right] I_0\left(\frac{\theta\phi}{2\sigma^2}\right) d\varphi,$$

and $\sigma$ is the beam width of the antenna and $I_0$ is the 0th order modified Bessel function. The angular correlation function, $C(\theta)$, is calculated as

$$C(\theta) = C(0) \left[1 + \frac{G(\theta)\theta^2}{\theta_{\rm c}^2}\right]^{-\frac{1}{2}} \exp\left[-\frac{\theta^2}{2(\theta_{\rm c}^2 + G(\theta)\theta^2)}\right], \tag{6}$$

where $G(\theta) = [\alpha_{\rm LS}(\theta) - 1]^2$. Here, $\theta_{\rm c}$ represents the coherence angle of the intrinsic temperature fluctuation determined by *e.g.* the Silk damping (Silk 1968). We use an interpolation formula: $G(\theta) = \theta_{\rm pr}^2/[\theta^2 + \theta_{\rm pr}^2 G(0)^{-1}]$. This formula well expresses the behavior of $G(\theta)$ in the entire region of $\theta$, as seen in figure 1. In figure 4, the average temperature anisotropy is plotted against $\theta$ for the cases of $\theta_{\rm c} = 4'$, $\sigma = 1'$, $z_{\rm F} = 2, 4$ and $10$. For the case of $z_{\rm F} =2$–$4$, the observed temperature anisotropy is smaller than the intrinsic one approximately by 40–60%.

We thank Tomoyoshi Ito, who developed GRAPE-2A, and Shaun Cole for many critical comments on the original manuscript. This research is partially supported by the Grand-in-Aid for Specially Promoted Research (04102002) of The Ministry of Education, Science,

**Reference**

Blanchard, A. and Schneider, J., 1987, *Astr. Astrophys.*, **184**, 1.

Cole, S. and Efstathiou, G., 1989, *Mon. Not. R. astr. Soc.*, **239**, 195.

Fukushige, T., Makino, J., Nishimura, O., and Ebisuzaki, T., 1994, submitted to *Publ. Astron. Soc. Japan*.

Gaier, T., Schuster, J., Gundersen, J., Koch, T., Seiffert, M., Meinhold, P., and Lubin, P., 1992, *Astrophys. J.*, **398**, L1.

Geller, M., and Huchra, J., 1989, *Science*, **246**, 897.

Goodman, J., Heggie, D. C. and Hut, P., 1993, *Astrophys. J.*, **415**, 715.

Gouda, N., Sasaki, M., and Suto, Y. *Astrophys. J.*, 1989, **341**, 557.

Gouda, N., and Sugiyama, N., 1992, *Astrophys. J.*, **395**, L59.

Gunn, J. E., 1967, *Astrophys. J.* **147**, 61.

Gurzadyan, V. G. and Savvidy, G. K., 1986, *Astr. Astrophys.*, **160**, 203.

Huang, S., Dubinski, J., and Carlberg, R. G., 1993, *Astrophys. J.* **404**, 73.

Ito, T., Makino, J., Fukushige, T., Ebisuzaki, T., Okumura, S. K., and Sugimoto, D., 1993, *Publ. Astron. Soc. Japan*, **45**, 339.

Kandrup, H. E. and Smith, H., 1991, *Astrophys. J.* **374**, 255.

Kandrup, H. E., Smith, H., and Willmes, D., 1992, *Astrophys. J.* **399**, 627.

Kashlinsky, A., 1988, *Astrophys. J.* **331**, L1.

King, I. R., 1966, *Astron. J.* **71**, 64.

Linder, E. V., 1990, *Mon. Not. R. astr. Soc.*, **243**, 353.

Lecar, M., 1968, *Bull. Astron.*, **3**, 91.

Makino, J. and Aarseth, S. J., 1992, *Publ. Astron. Soc. Japan.* **44**, 141.

Miller, R. H., 1964, *Astrophys. J.* **140**, 250.

Peebles, P. J. E. and Silk, J., 1990, *Nature*, **346**, 233.

Plummer, H. C., 1911, *Mon. Not. R. astr. Soc.*, **71**, 460.

Press, W. H., Flannery, B. P., Teukolsky, S. A. and Vetterling, W. T., 1986, *Numerical Recipes.* (Cambridge Univ. Press, London/New York).

Quinlan, G. D., and Tremain, S., 1992, *Mon. Not. R. astr. Soc.*, **259**, 505.

Readhead, A. C. S., Lawrence, C. R., Myers, S. T., Sargent, W. L. W., Hardebech, H. E. and Moffet, A. T., 1989, *Astrophys. J.* **346**, 566.

Sakagami, M. and Gouda, N., 1991, *Mon. Not. R. astr. Soc.* **249**, 241.

Sasaki, M. 1989, *Mon. Not. R. astr. Soc.*, **240**, 415.

Silk, J., 1968, *Astrophys. J.* **151**, 459.

Smoot, G. F., et al., 1992, *Astrophys. J.*, **396**, L1.

Suto, Y., 1991, *Publ. Astron. Soc. Japan.* **43**, L9.

Tomita, K. and Watanabe, K., 1989, *Prog. Theor. Phys.*, 1989, **82**, 563.

Uson, J. M. and Wilkinson, D. T., 1984, *Nature*, **312**, 427.

Watanabe, K. and Tomita, K., 1991, *Astrophys. J.*, **370**, 481.

Will, C. M., 1981, *Theory and experiment in gravitational physics.* (Cambridge Univ. Press, London/New York).

Wilson, M. L., and Silk, J., 1981, *Astrophys. J.* **243**, 14.

Table 1. The growth factor of angle between photons coming from the last scattering surface

| object | $r_{\rm h}$ ($h^{-1}$Mpc) | $d_{\rm sep}$ ($h^{-1}$Mpc) | $\Omega_{\rm s}$ | $d_{\rm pr}$ ($h^{-1}$Mpc) | $r_{\rm h}/d_{\rm pr}$ | $\theta_{\rm pr}$ (') | $\tau$ | $\eta$ | $\alpha_{\rm LS}$ [$z_{\rm F}=2$] | $\alpha_{\rm LS}$ [$z_{\rm F}=4$] |
|---|---|---|---|---|---|---|---|---|---|---|
| galaxy | 0.02 | 5 | 0.2 | 0.65 | 0.031 | 0.50 | 0.18 | 0.56 | 2.5 | 3.9 |
| cluster of gal. | 1 | 35 | 0.1 | 14 | 0.071 | 7.8 | 0.19 | 0.38 | 1.9 | 2.5 |
| supercluster | 5 | 150 | 1.0 | 71 | 0.070 | 122 | 0.19 | 1.20 | 7.3 | 18.3 |

**Figure Caption**

**Fig.1** Growth factor, $\alpha_{1.5} \equiv \theta_{1.5}/\theta_0$, plotted against $\theta_0/\theta_{\rm pr}$. The triangles, squares and pentagons are for $N = 128, 1024$, and $8192$, respectively.

**Fig.2** Growth factor, $\alpha_{1.5}$, plotted against $r_{\rm h}/d_{\rm pr}$ for different mass distribution of scattering objects. Solid curves indicate the King models. The triangles, squares and pentagons are for $W_0 = 5, 8$, and $12$, respectively. Dotted curve indicates the Plummer model.

**Fig.3** Growth of angle between photons coming from the last scattering surface. The redshift $z_{\rm LS}$ is that of the last (Thomson) scattering surface, $z_{\rm F}$ is the redshift when the scattering object were formed, $w$ and $\theta$ are the distance and angle between two rays, respectively.

**Fig.4** Averaged temperature fluctuation plotted against the chopping angle, $\theta$, for $\theta_c = 4'$ and $\sigma = 1'$. The thin curve indicates the intrinsic temperature fluctuation with no gravitational scattering. The thick curves indicate the observed temperature fluctuation for $z_{\rm F} = 2, 4$ and $10$.

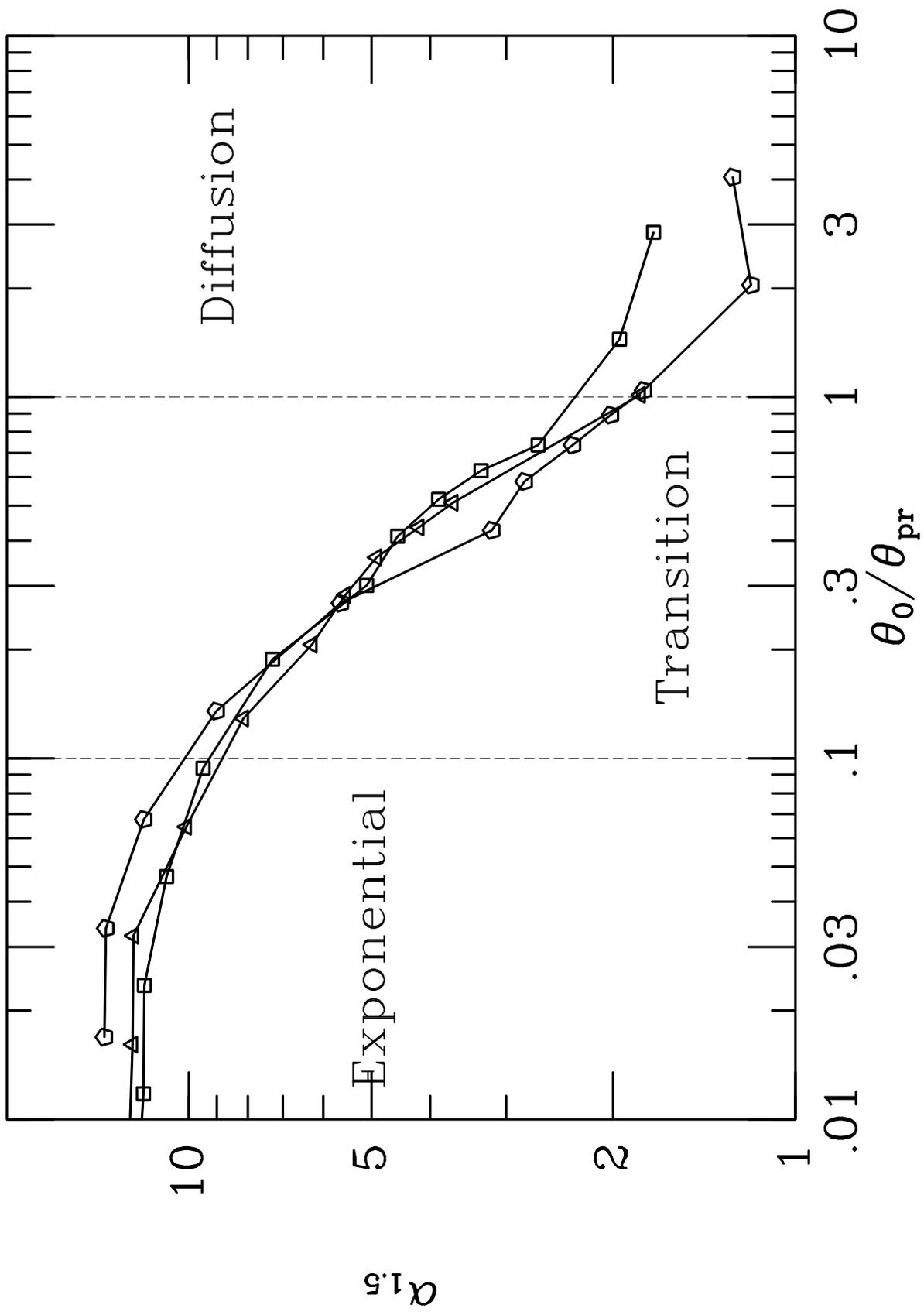

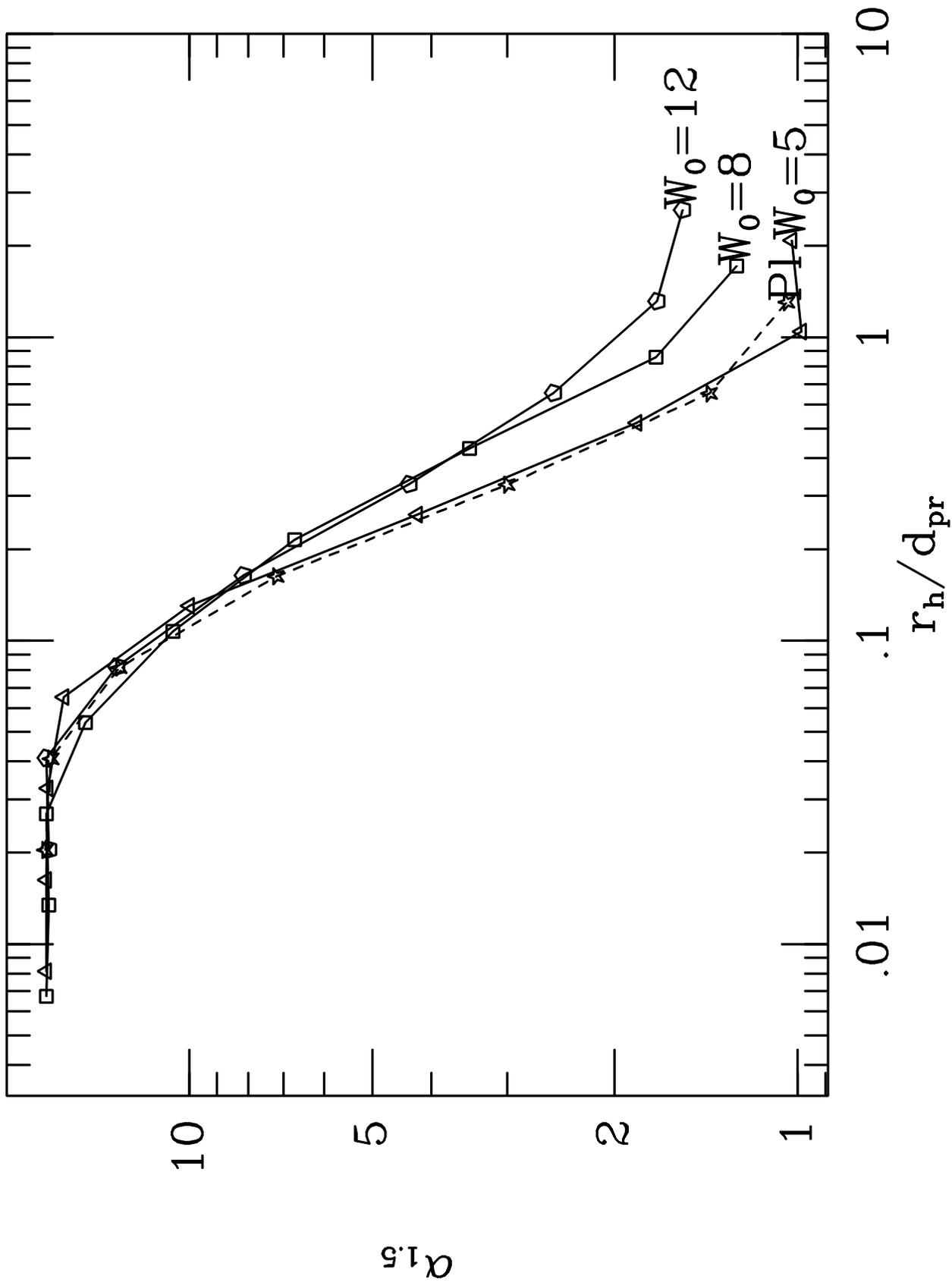

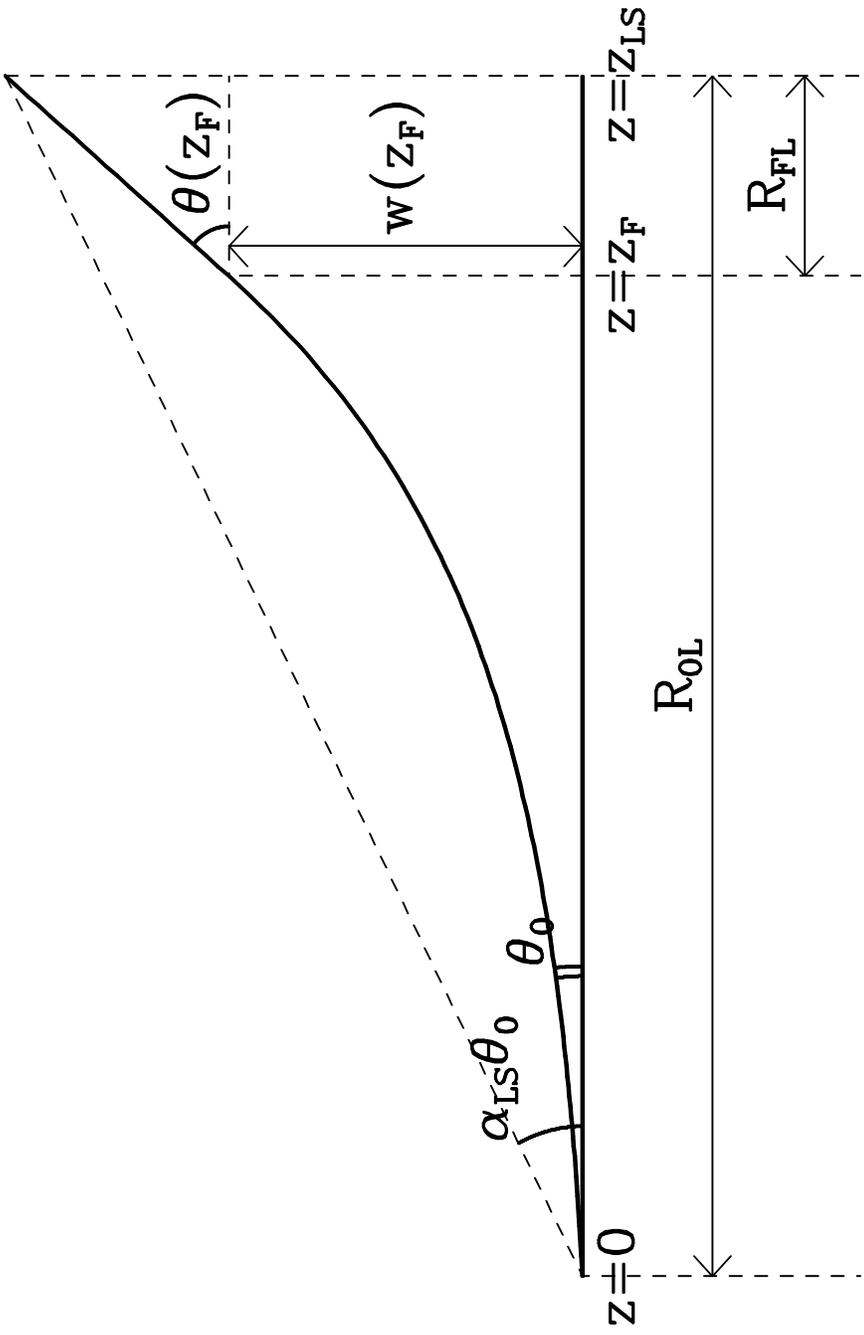

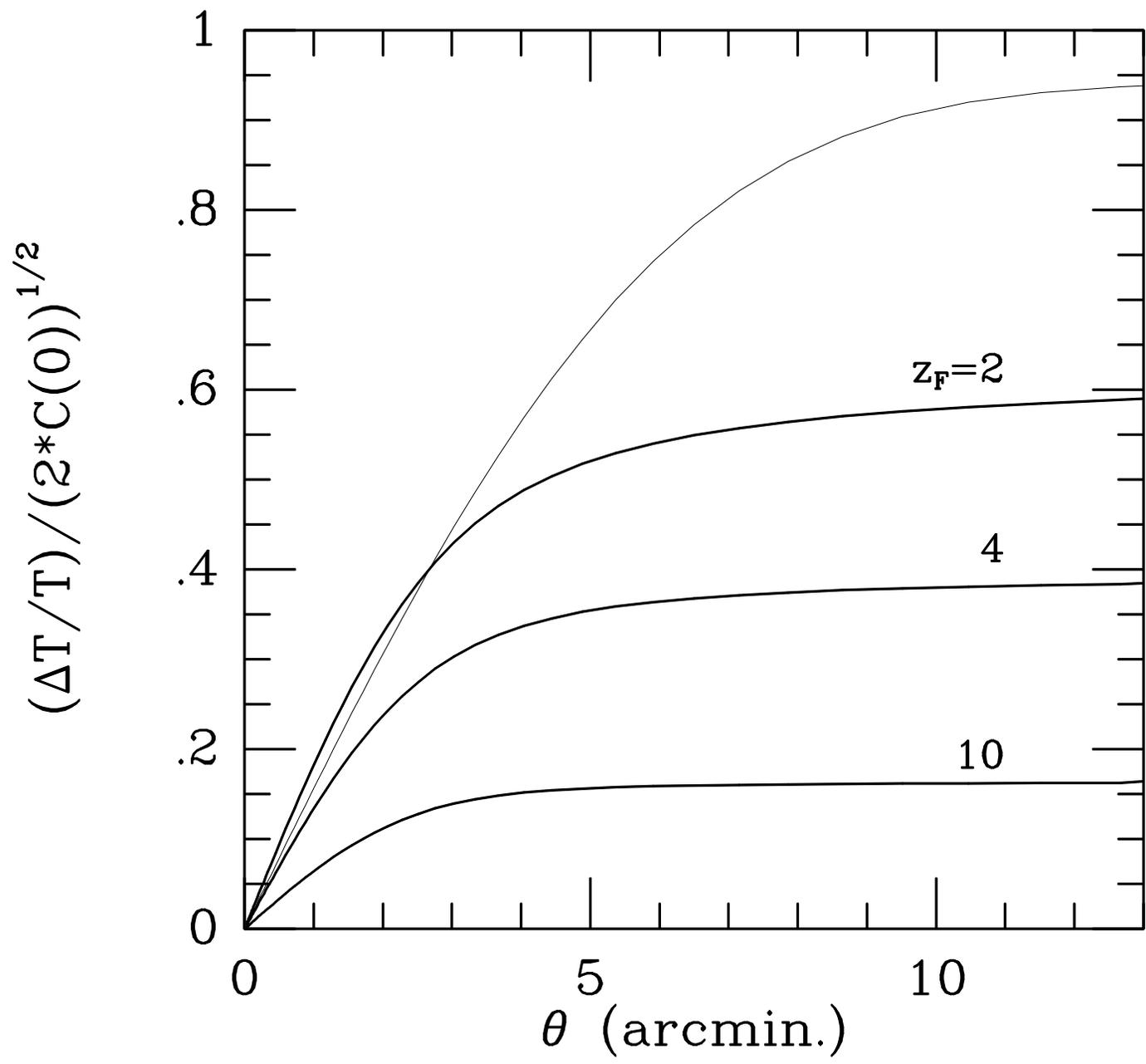